\documentclass[aps,prl,twocolumn,superscriptaddress,amsmath,showpacs,preprintnumbers]{revtex4}
\usepackage{epsfig}

\begin{document}
\preprint{\vtop{\hbox{RU06-06-B}
\vskip24pt}}

\title{\bf   Low Shear Viscosity due to Anderson Localization}

\author{Ioannis Giannakis}
\affiliation{Physics Department, The Rockefeller University,
1230 York Avenue, New York, NY 10021-6399}
\author{Defu Hou}
\affiliation{Institute of Particle Physics, Huazhong Normal
University, Wuhan, 430079, China} \affiliation{Physics Department,
The Rockefeller University, 1230 York Avenue, New York, NY
10021-6399}
\author{Jia-rong Li}
\affiliation{Institute of Particle Physics, Huazhong Normal University,
Wuhan, 430079, China}
\author{Hai-cang Ren}
\affiliation{Physics Department, The Rockefeller University,
1230 York Avenue, New York, NY 10021-6399}
\affiliation{Institute of Particle Physics, Huazhong Normal University,
Wuhan, 430079, China}
\begin{abstract}

 We study  the Anderson Localization effect on the shear viscosity in a system with random medium
 by Kubo formula. We show that this effect can reduce the
shear viscosity nonperturbatively. Then we discuss its possible implementation
 in heavy-ion collisions, where the created heavy bound states or other collective modes
may play the role of the random scatterer
  underlying Anderson Localization effect.
\end{abstract}

\maketitle

One of intriguing properties regarding the newly identified
quark-gluon plasma (QGP) created by relativistic heavy ion
collisions \cite{GM,star,phenix, Molnar} is its near perfect
hydrodynamical behavior. In particular the ratio of the shear
viscosity to the entropy density, deduced from the elliptic flow
measurement  is very small ${\eta}/{s}\simeq 0.2$ \cite{teaney}.
The weak coupling calculation gives a larger
 ratio of order $5.12(g^4\ln\frac{1}{g})^{-1}$ for $g<<1$
for three flavors of massless quark  \cite{yaffe, Baym} , with $g$
the QCD running coupling constant, which becomes unreliable near
$T_c$ where $g>1$.  Non-perturbative effects are expected  to play
an instrumental role to explain the observed ratio.

An analytical calculation of QCD beyond perturbation theory is not
available and the lattice simulation of transport coefficients
\cite{Nak} suffers from large error bars associated to the
analytic continuation to the mass shell, and so far only for
pure SU(3) Yang-Mills theory without quarks. A lower bound of the
viscosity-entropy ratio was estimated using the uncertainty
principle within the framework of kinetic theories $
\frac{\eta}{s}\ge \frac{1}{12}$\cite{dan}. This bound is
surprisingly close to that of the ${\cal N}=4$ supersymmetric
Yang-Mills theory at large $N_c$ and large 't Hooft coupling
following the conjectured AdS/CFT duality \cite{dts}
$\frac{\eta}{s}=\frac{1}{4\pi}\simeq 0.08$. The experimental
result is rather close to these bounds and the underlying QGP
seems strongly interacting\cite{LJL}.

Besides the lower bound of the viscosity-entropy ratio, there is
little understanding of the physical mechanism that contributes
to the suppression of the shear viscosity. One possibility is the
existence of many zero energy bound states as was suggested in
\cite{shuryak}. Another possibility is the anomalous viscosity in
a weakly but an expanding QGP within the formalism
of Boltzmann equation\cite{muller}.

In this letter, we shall explore a new mechanism that may lead
to a small viscosity in a system dominated by elastic scattering.
This is the Anderson Localization (AL) effect studied in the
context of condensed matter physics.

The prototype AL effect refers to the multiple scattering process
of a wave propagating in a disordered medium where the individual
scattering is elastic. Naively, one expects that the net scattered
wave equals to the incoherent sum of individual ones but a closer
look reals that this is not the case. The scattering amplitude
following any path of multiple scattering is in phase with the one
following its time reversed path when the scattering angle reaches
180, independent of the location of the individual scattering
events along the path. The elasticity or approximate one is
required to maintain the phase relation between the two scattering
waves. This coherence effect was first suggested by Anderson
\cite{anderson} and was verified experimentally by the intensity
peak of the reflected light off the surface of an amorphous
material with normal incidence \cite{mynard}(the figure 1 of this
paper provided an intuitive illustration of the AL effect). The
field theoretic treatments of AL have been developed in the
context of the electrical conductivity \cite{vol} in an amorphous
metal and the energy transport of light wave in a random medium
\cite{zbsu}.

In what follows, we shall generalize the existing analysis for the
Anderson Localization effect on the electrical conductivity and
the energy diffusivity to the momentum transport coefficient,
shear viscosity, by considering a massless scalar field in a
random medium. The massless particles resembles the thermal
partons in the QGP produced by RHIC. Then we assume that  there
are some random scatterers  required for the AL effect, for
instance, the heavy bound states produced in the heavy ion
collisions, or reminiscent of collective modes of the initial
state.  Because of the crudeness of our model, our conclusion
remains qualitative.

While the transport coefficients can be calculated by diagrammatic
expansion of the exact Kubo formula, they are usually extracted from
kinetic theory by the Chapman-Enskog approximation of the Boltzmann
equation. It has been shown that the latter approach
is equivalent to the ladder re-summation of the former one
for self interacting scalar field and pure Yang-Mills field
\cite{jeon1,hou}. The coherence effect underlying the Anderson
Localization corresponds to a set of maximally crossed diagrams
which is beyond the ladder ones, and therefore can not be obtained
from the kinetic theory.

Let us consider a simple model of massless scalar field in a random medium.
The Lagrangian reads
\begin{equation}
{\cal L}=-\frac{1}{2}\frac{\partial\phi}{\partial x_\mu}
\frac{\partial\phi}{\partial x_\mu}+\kappa\phi^2,
\end{equation}
where
\begin{equation}
\kappa=\sum_ju(\vec r-\vec R_j)
\label{potential}
\end{equation}
Here $u(\vec r-\vec R_j)$ is the potential of an impurity at the
random position  $R_j$ . The physics of AL is not sensitive to the
details of the coupling. The Kubo formula of the shear viscosity
is
\begin{equation}
\eta=\frac{1}{10}\lim_{\omega\to 0}\lim_{\vec q\to 0}{\rm Im}
\chi(\vec q,\omega)
\label{viscosity}
\end{equation}
where $\chi(\vec q, \omega)$ is the Fourier transformation of the
retarded two point Green function of the traceless stress tensor,
$\pi_{ij}=\frac{\partial\phi}{\partial
x_i}\frac{\partial\phi}{\partial x_j}
-\frac{1}{3}\delta_{ij}(\nabla\phi)^2$, i. e.
\begin{equation}
\label{kubo} \chi(\vec q,\omega)=\frac{i}{\omega}\int
d^4Xe^{-iQ\cdot X} <{\rm
Tr}{\bar\rho}[\pi_{ij}(X),\pi_{ij}(0)]>\theta(t).
\end{equation}
where  $\bar\rho=Ze^{-\beta H}$  is the the density operator with
 ${\rm Tr}{\bar \rho}=1$ . The bracket $<...>$ stands for the average over the centers of individual
impurity potentials, $\vec R_j$. The four vector notation
$X=(\vec r,it)$ for coordinate and $Q=(\vec q,i\omega)$ for momentum-energy are adapted.
The two point function  in eq.(\ref{kubo}) can be manipulated with the CTP( closed time
path ) formulation \cite{review} or by more traditional means \cite{vol}.
We find
\begin{eqnarray}
\label{ctpgf}
\chi(\vec q,\omega) &=& \frac{2}{\omega}\int\frac{d^3p}{(2\pi)^3}\int\frac{d^3p^\prime}{(2\pi)^3}
I_{ij}(\vec p,\vec q)I_{ij}(\vec p^\prime,-\vec q)\nonumber\\
&\times& \int_{-\infty}^{\infty}\frac{dp_0}{2\pi i}\{
[n\left(p_0^-\right)-n\left(p_0^+\right)]
\Phi^{RA}(P,P^\prime,Q)\nonumber\\
&-& n\left(p_0^-\right)\Phi^{RR}(P,P^\prime,Q)
+n\left(p_0^+\right)\Phi^{AA}(P,P^\prime,Q)\} \nonumber\\
&=& \chi^{RA}(\vec q,\omega)+\chi^{RR}(\vec q,\omega)+\chi^{AA}(\vec q,\omega)
\end{eqnarray}
where
$n(p_0)={(e^{\beta p_0}-1)}^{-1}$,
$p_0^{\pm}=p_0\pm\frac{1}{2}\omega$ and $I_{ij}(\vec p,\vec q)$ is
the Fourier component of the derivative operators in $\pi_{ij}$.
We have
\begin{equation}
I_{ij}(\vec p,0)I_{ij}(\vec p^\prime, 0)=\frac{2}{3}p^2p^{\prime 2}P_2(\hat p\cdot\hat p^\prime)
\label{vertex2}
\end{equation}
with $P_2(\hat p\cdot\hat p^\prime)$ the second Legendre polynomial.
\begin{figure}
\includegraphics[scale=0.48,clip=true]{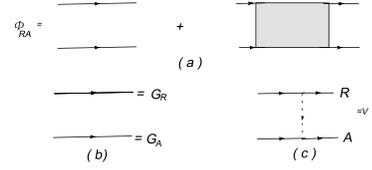}
\caption{\label{fig:epsart}. Diagrammatic representations  }
\end{figure}
The function $\Phi_{\alpha\beta}(P,P^\prime,Q)$ with $\alpha$ and
$\beta$ equal to R(retarded) or A(advanced) can be expressed in
terms the retarded and advanced full boson propagators
\begin{equation}
G_{R(A)}(P)=\frac{i}{p_0^2-p^2-\Sigma_{R(A)}(P)}
\end{equation}
together with
the 1PI vertex functions $\Gamma_{\alpha\beta}(P,P^\prime,Q)$, i.e.
\begin{eqnarray}
&&\Phi_{\alpha\beta}(P,P^{\prime},Q)=G_\alpha(P_+)G_\beta(P_-)\times
\\ \nonumber
&&\Big[(2\pi)^3\delta^3(\vec p-\vec p^{\, \prime})
-i\Gamma_{\alpha\beta}(P,P^\prime,Q)G_\alpha(P'_+)G_\beta(P'_-)\Big],
\label{phidef}
\end{eqnarray}
where $\Sigma_{R(A)}(P)$ is the retarded(advanced) self-energy
function, $P_{\pm}=P\pm\frac{1}{2}Q$, $P^\prime=(\vec
p^\prime,ip_0)$ and $P^{\prime\prime}=(\vec
p^{\prime\prime},ip_0)$. We have $\Sigma_R(P)=\rho\,
t_{\vec p,\vec p}(p_0)+O(\rho^2)$ where $t_{\vec
p^\prime,\vec p}(p_0)$ denotes the $T$-matrix of the scattering by
the potential $u(\vec r)$ and $\rho$ is the density of the
impurity centers. The Born approximation of $T$ matrix reads
$t_{\vec p^\prime,\vec p}(p_0)=u_{\vec p^\prime-\vec p}$ with
$u_{\vec p^\prime-\vec p}$ the Fourier transformation of $u(\vec
r)$. We have ${\rm Im}\Sigma_R(P)=-\rho p_0\sigma(p_0)+O(\rho^2)$ with
$\sigma(p_0)$ the cross-section of the scattering by a single term of
(\ref{potential}).

Since only $\Phi_{RA}(P,P^\prime,Q)$ contributes to the pinching
singularity of the $p_0$-integration at weak coupling, we shall
focus on it and see how the AL effect suppresses its contribution.
The subscript R, A of $\Gamma$ will be suppressed. The
diagrammatic representation of $\Phi_{RA}$ together with the
elements of the diagrams are displayed in Fig.1, where the
upper(lower) solid line stands for the retarded(advanced)
$\phi$-propagator and the dashed line denotes the impurity induced
interaction. The bare vertex of Fig.1c reads $V=-2i\pi
\rho\,t_{\vec p_+^\prime,\vec p_+}(p_0^+) t_{\vec p_-^\prime,\vec
p_-}^*(p_0^-)\delta(p_0^\prime-p_0)$, from which the full vertex
$\Gamma$ is generated. The energy delta function stems from the
elasticity of the underlying scattering, which leaves the energy
running along each solid line of $\Phi_{RA}$ conserved and all
loop integrals inside $\Phi_{RA}$ over spatial momenta only. It
follows from the Dyson-Schwinger equation for
$\Gamma(P,P^\prime,Q)$, shown in Fig.2, that
\begin{figure}
\includegraphics[scale=0.48,clip=true]{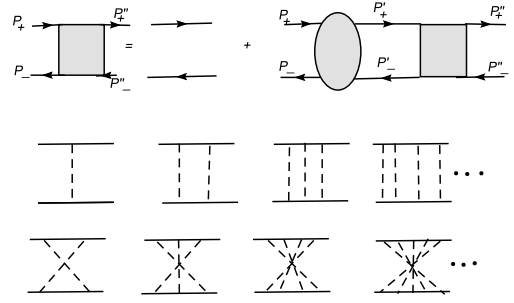}
\caption{\label{fig:epsart}. Schwinger Dyson equation and it's diagrammatic expansion}
\end{figure}
\begin{eqnarray}
&&\Gamma(P,P^\prime,Q)=\tilde \Gamma(P,P^\prime,Q) -\nonumber\\
&& i\int\frac{d^3 p''}{(2\pi)^3} \tilde \Gamma(P,P'',Q)G_R(P_+'') G_A (P_-'')
\Gamma(P'',P^\prime,Q) \label{DS}
\end{eqnarray}
which reduces to
\begin{eqnarray}
&&2(p_0\omega+2i \gamma p_0 -\vec p\cdot\vec q)f(P,P^\prime,Q)
=\tilde \Gamma(P,P^\prime,Q)\nonumber\\
&& -\int\frac{d^3 p''}{(2\pi)^3} \tilde \Gamma(P,P'',Q) \Delta G(P'',Q)
f(P'',P^\prime,Q) \label{DS}
\end{eqnarray}
where $f(P,P^\prime,Q)=\Gamma(P,P^\prime,Q)/[2(p_0\omega+2i \gamma p_0 -\vec
p\cdot\vec q)]$, $\Delta G(P,Q)=G_R(P_+)-G_A(P_-)$ and
$\tilde\Gamma(P,P^\prime,Q)$ is the 2PI part of
$\Gamma(P,P^\prime,Q)$. The leading order of $\tilde\Gamma$ reads
\begin{equation}
\tilde\Gamma_0=-i\rho \,t_{\vec p_+^\prime,\vec p_+}(p_0^+) t_{\vec
p_-^\prime,\vec p_-}^*(p_0^-).
\label{leading}
\end{equation}
For $|\Sigma_\alpha(P)|<<p$,
$\Delta G(P,Q)$ is sharply peaked at $p=p_0$ and we may
approximate $\Sigma_R(P)\simeq-\Sigma_A(P)\simeq -2ip_0\gamma$.
Introducing the partial wave expansion
\begin{equation}
\tilde\Gamma(P,P^\prime,Q)=\sum_l(2l+1)
c_l(p,p^\prime,\omega)P_l(\hat p\cdot\hat p^\prime)
\label{partial}
\end{equation}
at $Q=(0,\omega)$ and that of $f(P,P^\prime,Q)$,
an approximate solution to (\ref{DS}) at small $\omega$ and $\vec q$ can be
obtained
\begin{equation}
\Gamma(P,P^\prime,Q) =\frac{2i\gamma
c_0}{\omega+iDq^2}+4\pi\gamma\sum_{l=2}^\infty
\frac{(2l+1)c_l}{\gamma_l}P_l(\hat p\cdot\hat p^\prime),
\label{gammasol}
\end{equation}
where
\begin{equation}
\gamma_l
=\gamma+\frac{1}{4ip_0}\int\frac{d^3\vec p^\prime}{(2\pi)^3}
\tilde\Gamma(P,P^\prime,Q)P_l(\hat p\cdot\hat p^\prime)\Delta
G(P^\prime,Q) \label {gm}
\end{equation}
with the diffusion constant $ D(\omega)=\frac{1}{6\gamma_1}$ and the mass shell
approximation $\Delta G(P,Q)\simeq
2\pi\delta(p_0^2-p^2) {\rm sign}(p_0)$ has been employed.
The emergence of the hydrodynamical pole ( or the absence of $\gamma_0$ )
is the consequence of the Ward identity\cite{vol}
\begin{equation}
\Sigma_R(P)-\Sigma_A(P)=\int \frac{d^3\vec p_1^\prime}{(2\pi)^3}
\Delta G(P_1,Q)\tilde\Gamma(P_1,P,0).
\end{equation}

Inserting (\ref{gammasol}) into (8) and setting $Q=(0,\omega)$, we find
\begin{eqnarray}
\label{sol}
\Phi_{RA} &=& -\frac{\pi^2\Delta G(P,Q)
\delta(p-p^\prime)}{2\gamma p_0^3}\lbrace
1-\frac{c_0}{4\pi\omega}\nonumber\\
&&-\frac{\gamma}{8i\pi} \sum_{l\ge 1}\frac{(2l+1)P_l(\hat
p\cdot\hat p^\prime)}{\gamma_l}\rbrace
\end{eqnarray}
The shear viscosity picks up $l=2$ partial wave of $\Phi(P,P^\prime,Q)$.
It follows from  (\ref{viscosity}), (\ref{ctpgf}),
(\ref{vertex2}), (\ref{sol}) and the mass shell approximation that
\begin{eqnarray}
\label{chi}
\lim_{\omega\to 0}\chi_{RA}(0,\omega)
=\frac{i\beta}{6\pi^2}\int_0^\infty dp_0\frac{p_0^4 e^{\beta p_0}}
{\gamma_2(p_0)(e^{\beta p_0}-1)^2}.
\end{eqnarray}


The sum of the ladder diagrams shown in the 2nd line of Fig. 2, that corresponds
to the kinetic theory approach, can be obtained from the general
solution (\ref{gammasol}) with $\tilde\Gamma$ of (\ref{partial})
and (\ref{gm}) replaced by its leading order (\ref{leading}). The
corresponding $\gamma_l$'s will be denoted by $\gamma_l^{(0)}$. We
have $c_0=-8i\pi\gamma$ and
\begin{equation}
\label{1PI}
\Gamma_{\rm ladder}(P,P^\prime,Q)=\frac{16i\pi\gamma^2}{i\omega-D_0q^2}+...
\end{equation}
with the bare diffusion constant $D_0=\frac{1}{6\gamma_1^{(0)}}$. The result of the
shear viscosity following from the kinetic theory amounts to approximate $\gamma_2$
in the formula (\ref{chi}) by $\gamma_2^{(0)}$.

The AL effect induced by the backward coherent scattering is
reflected in the maximally crossed diagrams shown in the third
line of Fig. 2 which is a subset of 2PI diagrams for $\tilde\Gamma$.
The sum of this set of diagrams, $U(P,P^\prime,Q)$ can be obtained
from the that of ladder diagrams shown in the first line of Fig. 2
by reversing one of the rails ( dropping the bare vertex ). We
have
\begin{equation}
U(P,P^\prime,Q)=\Gamma_{\rm ladder}(P_{\rm rev},P_{\rm rev}^\prime,Q_{\rm rev})
\end{equation}
where $P_{\rm rev}=\Big[\frac{1}{2}(\vec p-\vec p^\prime+\vec q),ip_0\Big]$,
$P_{\rm rev}^\prime=\Big[\frac{1}{2}(\vec p^\prime-\vec p+\vec q),ip_0\Big]$
and $Q_{\rm rev}=(\vec p+\vec p^\prime,i\omega)$\cite{vol}. We find that
\begin{equation}
U(P,P^\prime,Q)=\frac{16i\pi\gamma^2}{i\omega-D_0(\vec p+\vec p^\prime)^2}
\end{equation}
and $\tilde\Gamma(P,P^\prime,Q)=\tilde\Gamma_0+U(P,P^\prime,Q)+...$.
Therefore the maximally crossed diagrams renders the integrand of (14)
to diverge in the backward direction, $\vec p^\prime\to-\vec p$, as
$\omega\to 0$, and this singularity extends to all partial waves.

In space dimensions $d<3$, the backward singularity is sufficient to
make $\gamma_l$
divergent at $\omega=0$ and thereby to make the dressed diffusion
constant vanish there for arbitrarily weak scattering. But a
critical strength of scattering is required for the purpose in
$d=3$. A self consistent treatment of the backward singularity by
summing up the most singular set of maximally crossed diagrams
amounts to replaced the bare diffusion constant in the expression
of $U(P,P^\prime,Q)$ with the dressed one, $D(\omega)$\cite{book}.
Eq. (\ref {gm}) for $l=1$ becomes then a self-consistent equation
for $D(\omega)$. Note that both $D_0$ and $D(\omega)$ depend on
the energy $p_0$. Upon introducing a cutoff $k_c$ that restricts
the momentum integration within the neighborhood of the backward
scattering and approximating $\Delta G(P,Q)$ by its peak
value\cite{footnote}, $\frac{1}{p_0 \gamma}$,
the self consistent equation for $D(\omega)$ becomes
\begin{equation}
\frac{1}{D}=\frac{1}{D_0}+\frac{24\pi\gamma^2}{p_0}
\int_{|\vec p+\vec p^\prime|<k_c}\frac{d^3\vec p^\prime}{(2\pi)^3}
\frac{\hat p\cdot\hat p^\prime}{i\omega-D(\vec p+\vec p^\prime)^2}.
\label{renorm}
\end{equation}
which is of the same form as that in cite{zbsu}.
In terms of the localization length defined in \cite{zbsu}, the limit
\begin{equation}
\xi=\lim_{\omega\to 0}\sqrt{i\frac{D}{\omega}},
\end{equation}
the onset of AL is characterized by $\xi\neq 0$, which implies
that $\lim_{\omega\to 0} D(\omega)=0$, and turns the hydrodynamical modes
into localized ones via $i\omega-D(\vec p+\vec p^\prime)^2\to
i\omega[1+(\vec p+\vec p^\prime)^2/\xi^2]$. Carrying out the integration of
(\ref{renorm}), we obtain the equation for the localization
length,
\begin{equation}
k_c\xi-\tan^{-1}(k_c\xi)=\frac{\pi p_0^2\xi}{12\gamma}.
\end{equation}
A real and nonzero solution for $\xi$ exists for
\begin{equation}
\frac{\pi p_0^2}{12\gamma^2}<1,
\label{broad}
\end{equation}
where we have set the cutoff $k_c\sim\gamma$. It follows from
eq.(\ref {gm}) for $l=2$ ( with $D_0$ replaced by $D$ in $\Gamma$
) that AL effect renders $\gamma_2(p_0)\to\infty$ when the
condition (\ref{broad}) is met. While it is unlikely that the
condition (\ref{broad}) holds for all $p_0$ ( because of the
unitary bound that prevents $\gamma(p_0)$ from growing indefinitely with
$p_0$ ), a finite domain of $p_0$ where AL takes place is quite
possible. If the AL domain covers the peak of the function
$p_0^4e^{\beta p_0} /(e^{\beta p_0}-1)^2$ in the integrad of the
formula (\ref{chi}), a significant reduction of the shear
viscosity is expected. In the parallel case of the Mie scattering of
an electromagnetic wave by a random ensemble of metallic spheres
studied in \cite{zbsu}, AL does occur for $p_0$ greater than a critical
value. But the upper bound of the AL domain was not reported
there.
But we have to admit that the
mass-shell approximation employed to obtain the approximate
solution (\ref{sol}) becomes marginal for $d=3$ because of the
condition (\ref{broad}).

The self-consistent equation parallel to (\ref{renorm}) for $d=2$
reveals that the localization length survives for an arbitrarily
week coupling and for all $p_0$. No further conditions like (\ref{broad})
are required.

In conclusion, we have proposed a non-perturbative mechanism, the
Anderson Localization effect, that may suppresses considerably the
shear viscosity and other transport coefficients. The AL effect is
caused by multiple coherent scattering in the system, which goes
beyond the Boltzmann equation. It can be treated
systematically in $d=2$ and self-consistently in $d=3$. Although
the AL effect is discussed in a simple toy model in this letter,
the physics of AL is more general. It only requires the individual
scattering event to be sufficiently elastic. Regarding the QGP
produced in RHIC, a potential source for AL is the scattering by
the heavy bound states produced in collisions, whose masses could
be as heavy as 2GeV. If such states behaves like NR particles, the
momentum transfer in a scattering event could exceed by far the
energy transfer. Other non-equilibrium effect, such as the
color gauge field produced in an expanding QGP \cite{muller} may also
implement the elastic scattering necessary for AL. Much study is still
needed to ascertain the relevance of AL to the low viscosity of sQGP and
other RHIC phenomenology, say the energy losses process.

\begin{acknowledgments}

We would like to thank Professors A. Libchaber and Z. B. Su from
whom we learnd the physics of Anderson Localization.  We thank R.
Pisarski, D. Rischke and T. Sch\" afer for interesting discussions
. The work of I. G., H. C. R and D. F. H is supported in part by
US Department of Energy under grants DE-FG02-91ER40651-TASKB. The
work of D. F. H.  and H.  C.  R. is supported in part by NSFC
under grant No.  10575043. The work of D. F. H.  is also
supported in part by  Educational Committee under grants
NCET-05-0675 and 704035.

\end{acknowledgments}



\begin{thebibliography}{99}
\bibitem{footnote} The mass shell approximation has to be relaxed here because its application would
lead to an artificial divergence of the integral over $\vec p^\prime$.
\end{thebibliography}
\end{document}